\documentclass[letterpaper]{article}
\usepackage[utf8]{inputenc}

\usepackage{amsmath}
\usepackage{amsfonts}
\usepackage{amssymb}
\usepackage{amsthm}
\usepackage{authblk}
\usepackage{booktabs}
\usepackage{cite}
\usepackage{enumerate}
\usepackage{graphicx}
\usepackage[left=2cm,right=2cm,top=2cm,bottom=2cm]{geometry}
\usepackage{mathabx}
\usepackage{multirow}
\usepackage{subcaption}

\usepackage{color}

\author[1,2]{Pasquale Bosso\thanks{pasquale.bosso@uleth.ca}}
\author[1]{Saurya Das\thanks{saurya.das@uleth.ca}}
\affil[1]{Theoretical Physics Group and Quantum Alberta, 
	\protect\\ 
	Department of Physics and Astronomy, 
	\protect\\ 
	University of Lethbridge,
	\protect\\ 
	4401 University Drive, Lethbridge, Alberta, Canada, T1K 3M4\vspace{1em}}
\affil[2]{Divisi\'on de Ciencias e Ingenier\'ias, Universidad de Guanajuato,\protect\\ Loma del Bosque 103, Lomas del Campestre CP 37150, Le\'on, Gto., M\'exico}

\title{Comments on ``Schwinger's Model of Angular Momentum with GUP'' by H. Verma et al, arXiv:1808.00766}

\date{}

\begin{document}
	
	\maketitle
	
	\begin{abstract}
		In this note, we show that the methodology and conclusions of ``Schwinger's Model of Angular Momentum with GUP'' 
		[arxiv:1808.00766] are flawed and that the conclusions of "Generalized Uncertainty Principle and angular momentum" (P. Bosso and S. Das) 
		[arxiv:1607.01083] remain valid.
	\end{abstract}
	
	\section{Introduction}
	In \cite{Bosso2017}, we studied the modification of the angular momentum algebra dictated 
	by a modification of the commutation relations between position and momentum.
	This led to interesting results concerning the angular momentum spectrum, energy spectrum of the hydrogen atom, 
	interactions between magnetic fields and angular momenta, and a description of multi-particle systems, 
	including corrections to Clebsch-Gordan coefficients.
	Among the noteworthy results of that paper, we have that the commutation relations for angular momentum operators now involve also 
	functions of the linear momentum operator.
	However, following usual steps, we showed that the spectrum involves only expectation values of this function.
	Notice that the appearance of expectation values at this level is not the result of any approximation or assumption.
	On the other hand, when the radial Schr\"odinger equation for the hydrogen atom is studied, 
	the assumption that the modification appears only via its expectation value is considered.
	This is clearly an approximation, useful in that the equation resemble that of standard quantum mechanics with corrections that have implication in the emission/absorption wavelengths.
	It is important to notice that this approximation is considered only in the study of the hydrogen atom and of the interaction between magnetic fields and angular momenta.
	The rest of \cite{Bosso2017} does not rely on this assumption.
	
	That paper concludes with an analysis of the theory of angular momentum for systems of many particles.
	Specifically, we computed some of the Clebsch-Gordan coefficients.
	We showed that the Clebsh-Gordan coefficients are modified due to GUP. 
	However, we also showed that the modification is associated with an ambiguity: 
	coefficients for the same system but computed starting from higher or lower angular momentum 
	states are in general different.
	This is something to expect, because the spectrum of the angular momentum operators, 
	as well as the action of the ladder operators $J_-$ and $J_+$, is influenced by the linear 
	momentum of the states to which they are applied.
	This ambiguity is a direct consequence of the modified position-momentum uncertainty relation considered in \cite{Bosso2017} and needs further study.
	
	\section{Comments on \cite{Verma2018}}
	
	Verma et al., in \cite{Verma2018}, have tried a possible resolution of the above ambiguity.
	Below, we comment on the main aspects of their paper.
	%
	\paragraph{Ladder operators for a simple harmonic oscillator (SHO):}
	
	The authors define a couple of operators, $A$ and $A^\dagger$, in terms of the position and momentum operators as follows
	\begin{align}
	A = & \sqrt{\frac{2 m \omega}{\hbar}} \left(q + \frac{i}{m \omega} p\right), &
	A^\dagger = & \sqrt{\frac{2 m \omega}{\hbar}} \left(q - \frac{i}{m \omega} p\right).
	\end{align} 
	As the authors themselves show, these operators are not ladder operators for a SHO since $[A,A^\dagger] \neq 1$.
	To avoid this problem, they divide each operator by the square root of the expectation value of $[A,A^\dagger]$.
	This strategy can only be considered as an approximation, not as a rigorous method to obtain ladder operators.
	Furthermore, this approximation leads to trivial results, effectively accounting for a scaling of the Planck constant.
	Finally, it is not necessary.
	In fact, it is always possible to construct a new set of operators that act as ladder operators for the GUP-modified
	system under study, as shown in \cite{Bossotilde} (see \cite{Bosso2017a} for an application).
	
	\paragraph{Substituting the modification $\mathcal{C}$ by its expectation value:}
	The authors motivate its use by saying that it was previously used in our work \cite{Bosso2017}.
	However, as we explained above, this is not entirely true.
	This substitution was in fact used to obtain GUP-modifications for some specific examples, namely the hydrogen atom and the magnetic field-angular momentum interaction, and which can potentially be verified experimentally.
	However, the rest of the results in \cite{Bosso2017} are not based on such substitution.
	In particular, the commutation relations between angular momentum operators are computed implementing directly the modified commutation relation between position and momentum operators.
	Furthermore, the appearance of the expectation value $\langle \mathcal{C} \rangle$ in the angular momentum spectrum was derived from rigorous standard procedures.
	Nonetheless, this does not motivate in general an indiscriminate substitution $\mathcal{C} \rightarrow \langle \mathcal{C} \rangle$.
	
	\paragraph{The Jordan map and Schwinger's model for the angular momentum:}
	The Jordan map connects a set of matrices to a set of harmonic oscillator operators \cite{Jordan1935}.
	In the specific case of the angular momentum, to each matrix representing an angular momentum operator, with elements $J_{ij}$, the Jordan map associates an operator $J = \sum_{i,j} a^\dagger_i J_{ij} a_j$.
	Therefore, in the standard theory of quantum mechanics, we find the following expressions \cite{biedenharn1981angular}
	\begin{subequations}
		\begin{align}
		J_x = & \frac{\hbar}{2} (a_1^\dagger a_2 + a_2^\dagger a_1),&
		J_y = & i \frac{\hbar}{2} (a_2^\dagger a_1 - a_1^\dagger a_2),&
		J_z = & \frac{\hbar}{2} (N_1 - N_2),\\
		J^2 = & \frac{\hbar^2}{4} (N_1 + N_2)(N_1 + N_2 + 2),&
		J_+ = & J_x + i J_y = \hbar a_2^\dagger a_1,&
		J_- = & J_x - i J_y = \hbar a_1^\dagger a_2.
		\end{align}
	\end{subequations}
	Notice that the expressions, used by Verma et al. in \cite{Verma2018}, are constructed from the representation of the angular momentum operators as Pauli matrices.
	A property of the Jordan map is that it preserves commutation relations.
	In other words, whatever the commutation relations for the angular momentum matrices are, the operators in terms of ladder operators will have the same relations, since they are inherited from the matrices.
	Notice therefore that
	\begin{enumerate}[(i)]
		\item the procedure used in \cite{Verma2018} is flawed, because their $A, A^\dagger$ operators are
		{\it not} ladder operators, as explained above,
		\item tautological, as starting from a representation of the standard (non GUP) angular momentum algebra, the authors use the Jordan map to find the commutation relations for the angular momentum operators, which necessarily are the standard one.
	\end{enumerate}
	
	As a final comment, it is worth noticing that their results concerning the commutation relations of the angular momentum operators do not match the results in \cite{Bosso2017}, obtained via direct computations.
	
	\paragraph{Clebsch-Gordan coefficients for a system of two particles with GUP:}
	Finally, we comment on the computation of the Clebsch-Gordan coefficients.
	The authors compute the coefficients starting from a maximal angular momentum state and applying the operator $J_- = j_{1,-} + j_{2,-}$.
	They proceed by imposing normalization conditions.
	
	It is important at this point to focus on a particular aspect.
	Consider the tower of states derived by repeated applications of $J_-$ on the top state $|L,L\rangle$.
	Each of these states has to be normalized independently from the others.
	On the other hand, each of these states can be in general represented in multiple ways in terms of the constituent states.
	More precisely, given a total angular momentum state $|L,M\rangle$, this describes any composition of two systems, with angular momentum states $|l_1,m_1\rangle$ and $|l_2,m_2\rangle$, such that $|l_1 - l_2| \leq L \leq l_1 + l_2$ and $M = m_1 + m_2$.
	In particular, for every value $M$, in general several configurations will be possible such to obtain different values of $L$.
	Every choice of these compositions is described by a different Clebsch-Gordan coefficient.
	
	Having this picture in mind, it is clear that the normalization of Clebsch-Gordan coefficients has to be performed considering the independent normalization of states $|L,M\rangle$ with different values $M$ and the orthogonality among the states of equal $M$ but different $L$.
	However, in \cite{Verma2018} the authors seem to normalize states belonging to different values of $M$ simultaneously, resulting in constraints such as 
	\begin{equation}
		\langle \mathcal{C}_1\rangle = \langle \mathcal{C}_2 \rangle = \langle \mathcal{C} \rangle, \qquad \mbox{with } \mathcal{C}_i = 2 \alpha p_i - 4 \alpha^2 p^2_i,
	\end{equation}
	where $\mathcal{C}_1$  and $\mathcal{C}_2$ represent the modifications associated with the two constituent systems, while $\mathcal{C}$ is the modification associated with the entire system.
	Such a relation is clearly unacceptable, since it implies conditions on the linear momenta of a composite system and its constituents.
	Since the constituents' momenta are independent physical quantities, no such conditions can exist.
	Furthermore, the only condition on the composite momenta is that it is the vector sum of the constituent momenta.
	
	\vspace{1em}
	
	The results of \cite{Bosso2017}, therefore, still stand, and the objections raised in \cite{Verma2018} are invalid. 
	Nevertheless, it is possible that a correct application of the Jordan map to the GUP-modified angular momentum operators 
	may be an useful tool in gaining insight into the problem in question and its possible resolution.
	

	\vspace{1cm}
	\noindent
	{\bf Acknowledgment}
	
	\vspace{0.1cm}
	\noindent
	We would like to thank the authors of \cite{Verma2018} for correspondence.

\end{document}